\documentstyle[11pt,newpasp,twoside,epsf]{article}
\def\edcomment#1{\iffalse\marginpar{\raggedright\sl#1\/}\else\relax\fi}
\reversemarginpar

\begin{document}
\title{The FIR/Radio correlation of high redshift glaxies in the region
  of the HDF-N} 
\author{M.A. Garrett (garrett@jive.nl)} 
\affil{Joint Institute for VLBI in Europe, Postbus 2, 7990~AA Dwingeloo, NL.}

\begin{abstract}
  
  The correlation between the far-infrared (FIR) and radio emission is
  well established for nearby star forming galaxies. Many applications,
  in particular the radio-to-submm spectral index redshift indicator,
  tacitly assume that the relation holds well beyond our local
  neighbourhood, to systems located at cosmological distances. In order
  to test this assumption I have constructed a sample of 22 HDF-N
  galaxies, all with measured spectroscopic redshifts, and all detected
  by {\it both} ISO and the WSRT at 15~micron and 1.4~GHz respectively.
  The galaxies span a wide range of redshift with a median value of $z
  \sim 0.7$. The ISO 15 micron data were k-corrected and extrapolated
  to the FIR (60 and 100 micron) by assuming a starburst (M82) spectral
  energy distribution (SED) for the entire sample. An initial analysis
  of the data suggests that the correlation between the FIR and the
  radio emission continues to apply at high redshift with no
  obvious indication that it fails to apply beyond $z \sim 1.3$.  The
  sample is ``contaminated'' by at least 1 distant ($z=4.4$), radio-loud AGN,
  VLA~J123642+621331. This source has recently been detected by the
  first deep field VLBI observations of the HDF-N and is clearly
  identified as an ``outlier'' in the FIR/radio correlation.  I
  briefly comment on the impact upgraded and next generation radio
  instruments (such as $e$-MERLIN and the Square Km Array) can have in
  studies of star formation in the early Universe.

\end{abstract}

\section{Introduction}

The correlation between the far-infrared (FIR) and radio emission is
one of the tightest and most universal correlations known among the
global properties of {\it local} star forming and starburst galaxies
(see Helou \& Bicay 1993, and references therein). Entirely unexpected,
yet extending across five orders of magnitude in luminosity, the
physical explanation for the tightness of the relation is that the
non-thermal radio emission and the thermal FIR emission are both
related to processes governed by massive star formation.  Established
via FIR and radio observations of {\it nearby} galaxies, the FIR/radio
correlation forms the basis of several key applications. In
particular, the technique of using the radio-to-submm spectral index as
a redshift indicator (Carilli \& Yun, 2000) and the use of unbiased
radio observations to estimate the global star formation history of the
Universe (Haarsma et al.  2000). These applications are beginning to
completely overhaul our ideas of star formation and galaxy formation in
the early Universe. In particular, deep SCUBA sub-mm observations of
the high redshift Universe (e.g. Hughes et al. 1998) are beginning
to reveal a dominant population of dusty, optically faint (early type)
galaxies that are inferred (via the radio-to-submm spectral index
redshift indicator) to lie at cosmological distances ($z>2$).

A basic, yet crucial assumption that underpins these developments is
that the FIR/radio correlation is entirely independent of redshift.
However, there are many reasons why the correlation might well fail at
non-local redshifts. In the radio domain these include (Condon 1992,
Lisenfeld, Volk \& Xu 1996), the quenching of the radio emission due to
inverse Compton (IC) losses via the CMB (scaling as $(1+z)^{4}$). In
addition, the trend for higher redshift systems to be more luminous (a
selection effect related to current sensitivity levels) may also lead
to global changes in the properties of the detected sources (as
compared to local, less luminous star forming galaxies that form the
basis of the locally derived relation). An overall change in the SED of
these systems might well be expected, and the differing time scales
associated with the rise in the FIR and radio emission may be
significant, particularly for vigorous starburst systems.  Specific
effects regarding the level of radio emission might occur via (i)
evolving magnetic field strengths (Dunne et al. 2000), (ii) varying
levels of free-free absorption (Rengarajan \& Takeuchi 2001), (iii)
further IC losses associated with the intense {\it local} radiation
environment and (iv) the possibility of significant ``contamination''
of high redshift star forming systems with co-existing low-luminosity,
``radio loud'' AGN.

Until recently a study of the FIR/radio correlation at non-local
redshifts has been difficult since it requires extremely deep and
complementary observations in both the radio and far/mid-IR wavebands,
together with spectroscopic redshifts of relatively faint sources. Like
many other areas of astrophysics, the situation has recently been
transformed by the wealth of publicly available data generated by
deep multi-wavelength studies of the Hubble Deep Field (HDF). In this
paper I investigate the nature of the FIR/radio correlation at moderate
redshifts (up to $z\sim 1.3)$ assuming throughout the currently
``preferred'' cosmological model ($\Omega_{m}=0.3$,
$\Omega_{\Lambda}=0.7$, $H_{0}=70$~km/sec/Mpc).

\section{The ISO 15 micron and WSRT 1.4~GHz HDF-N Sample} 

A sample of 22 mostly low and moderate redshift HDF-N sources was
established from the following simple criteria: (i) they are detected
by {\it both} the WSRT at 1.4~GHz (Garrett et al. 2000) and ISO at 15
micron (Aussel et al. 1999) and (ii) each source is independently
identified with the same optical candidate with a measured redshift
(e.g. Cohen et al.  2001). No attempt was made to remove AGN candidates from
the sample.

One limitation of this study is that the deep ISO measurements were
made in the mid-infrared and it was therefore necessary to extrapolate
the 15 micron measurements to 60 and 100 micron. This was achieved by
constructing a SED based on the starburst galaxy M82 (this source has
been extensively studied at all wave-bands including those that form the
basis of our SED - the radio, mm, sub-mm, far, mid and near infrared
bands).  A k-correction (dependent on source redshift and our assumed
SED) was also applied to both the mid-infrared and radio data. Note
that for non-local redshifts the steepness of the Rayleigh-Jeans tail
in the FIR makes the k-correction absolutely essential, without this
the {\it observed} FIR-Radio correlation completely disappears (it is
this same property that, in principle, makes the sub-mm/radio spectral
index such a powerful redshift estimator). I adopt the FIR/radio ratio
as quantified in the ``q'' parameter defined by Condon (1992) viz:
$q=log(\frac{S_{100}+2.6S_{60}}{3S_{1.4}})$

\begin{figure}
\plotfiddle{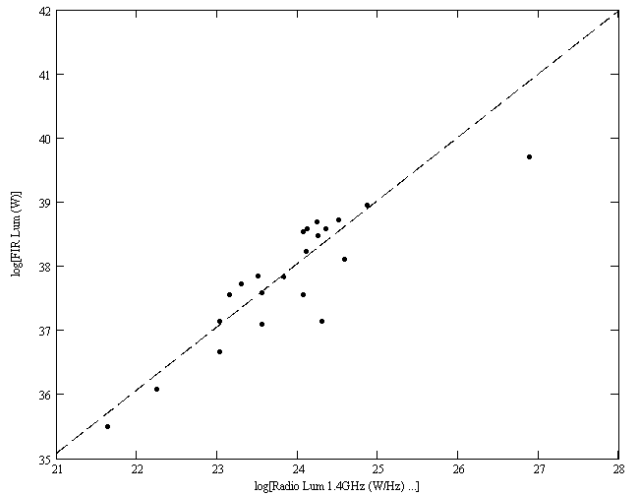}{7cm}{0}{37}{37}{-130}{5} 
\caption{A logarithmic plot of the FIR vs Radio luminosity of the
  sample. A median fit to the data (solid line) is also presented - 
  the FIR/radio correlation is clearly seen to apply out to
  $z\sim1.3$. The distant ($z=4.4$) radio-loud AGN 
  shows up as an outlier at the far RHS of the plot.}
\end{figure}

\section{Results} 

Figure~1 shows a logarithmic plot of the FIR and radio luminosities for
the HDF-N sample. A median-fit to the data indicates the linearity of
the relation (the slope of the fit is 0.99 with a correlation
coefficient of 0.89). This striking result strongly suggests that the
FIR/radio correlation continues to apply at non-local, moderate
redshifts.  Note that the luminosities probed by our faint (but
distant) sample extends the range (upwards) by over an order of
magnitude compared to previous, ``local'' studies. The highest
luminosity sources in Figure~1 are also the most distant.

Figure~2 shows the value of {\it q} plotted as a function of redshift.
There is a clear clustering of sources around $q\sim 2$ (excluding the
radio-loud outliers). Given the dominant error involved in the process of
extrapolating the mid-infrared fluxes to the far-infrared, this is
similar to the value measured by Condon (1992) for local galaxies -
$q\sim 2.3$. 

There are a few outliers observed in both Figure~1 and 2. One of these
(the data point lying to the extreme right of Figure~2) is also
(somewhat alarmingly) the highest redshift source in the sample.
However, this $z=4.4$ source (VLA~J123642+621331) is widely believed to
be an optically faint, dust obscured star forming galaxy, which also
harbours an AGN (Waddington et al. 1999, Garrett et al. 2001, Brandt et
al.  2001). The fact that it deviates from the FIR/radio correlation is
in fact rather reassuring, and suggests that just as the locally
derived correlation can distinguish between sources powered by AGN and
star formation processes, the extended correlation can very likely do
the same thing for moderate redshift sources. Figure~3 shows the
European VLBI detections of three AGN in the HDF-N region (including
our radio-loud AGN candidate, VLA~J123642+621331).  The EVN detection
implies a small physical size for the radio emitting region and this,
together with the radio source luminosity, strongly suggests that a
radio loud AGN dominates the total radio luminosity of the source. Note
that extremely deep, VLBI observations of the dust obscured, SCUBA
source population, may be the best (only) way of currently
disentangling the relative contribution made by AGN and star formation
processes in these systems.

\begin{figure}
\plotfiddle{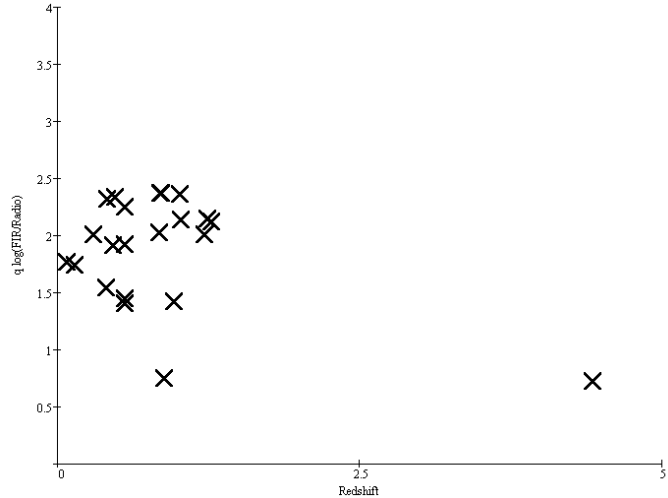}{7cm}{0}{37}{37}{-130}{5} 
\caption{The ratio, $q$ of the FIR/Radio emission plotted against
  redshift for each galaxy in the sample. Radio-loud AGN clearly
  show-up as outliers on the graph.}
\end{figure}

\begin{figure}
\plotfiddle{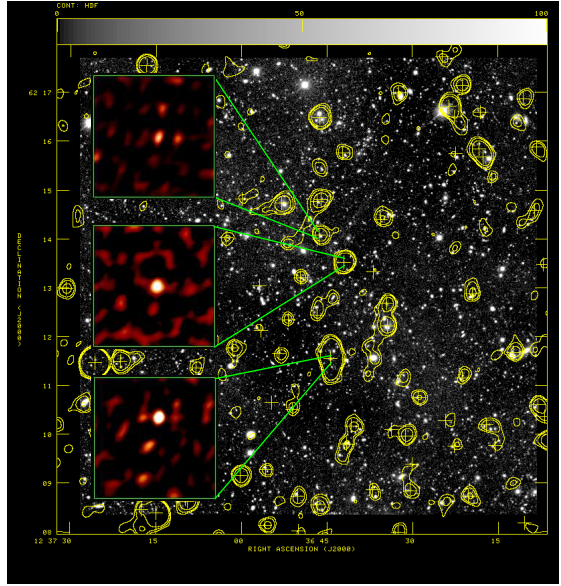}{8cm}{0}{27}{27}{-115}{-41} 
\caption{The three sources (including VLA~J123642+621331 - middle insert) 
  simultaneously detected by deep, wide-field EVN observations of the
  HDF-N.}
\end{figure}

\begin{figure}
\plotfiddle{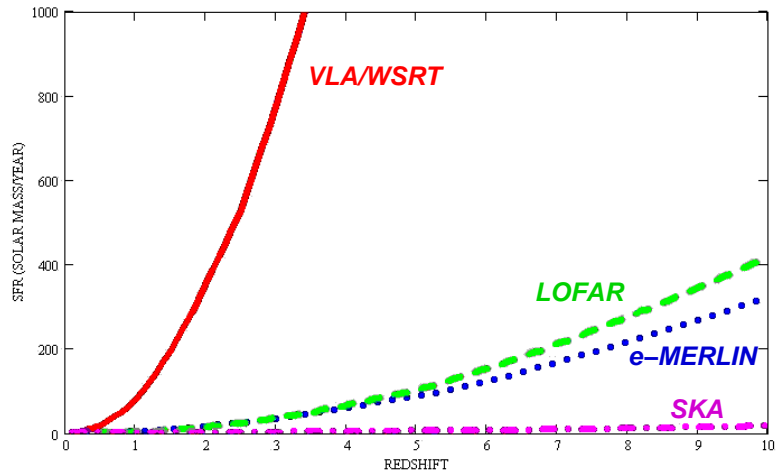}{7cm}{0}{49}{49}{-150}{4}
\caption{The (3 sigma) Star formation rates that can be probed by 
  (i) deep field studies using current instruments e.g. 1.4~GHz WSRT \& VLA -
  48 hour integration (solid line), (ii) an 8-beam 120 ~MHz LOFAR - 1 month
  integration (dashed line), (iii) an upgraded 5~GHz $e$-MERLIN or EVLA - 96
  hour integration (dotted line) and (iv) the 1.4~GHz Square Km Array - 12 hour
  integration (dot-dash line).}
\end{figure}

\section{Conclusions} 

The FIR/radio correlation continues to delight - in addition to the
puzzle of just why the correlation is so tight in the local Unverse, it
now appears we must also consider why it continues to apply to the more
distant starburst galaxies presented here. There is no obvious evidence
to suggest that the relation does not apply equally well beyond $z\sim
1$. This is particularly happy news for techniques (such as the Carilli
\& Yun (2000) radio-to-submm spectral index redshift indicator) that
implicitly assume this to be the case.  The result also serves to
re-emphasise the crucial role deep radio observations play as an
unbiased estimator in studies of the star-formation history of the
(largely dust obscured) early Universe. A more extensive investigation
of the FIR/radio correlation at high redshift awaits much deeper SIRTF
and ALMA infrared and sub-mm observations. Complimentary radio data
will be essential and these depend critically on upgraded instruments
such as those currently proposed ($e$-MERLIN and EVLA). However, truly
next generation radio instruments, in particular the Square Km Array
(SKA), will {\it completely} revolutionise our view of star formation
in the Universe.  SKA represents a quantum leap in our capability of
detecting even relatively feeble/normal ($\sim$~few~$M_{\odot}$/yr)
star forming galaxies out to any redshift that they might reasonably be
expected to exist. Figure~4 shows the star formation rate than can be
probed by current, upgraded and next generation radio instruments,
assuming typical (but different) integration times.

\acknowledgements

The Westerbork Synthesis Radio Telescope is operated by the ASTRON
(Netherlands Foundation for Research in Astronomy) with support from
the Netherlands Foundation for Scientific Research, NWO.  The European
VLBI Network is a joint facility of European and Chinese radio
astronomy institutes funded by their national research councils.

\end{document}